# Designing the Architecture of a Convolutional Neural Network Automatically for Diabetic Retinopathy Diagnosis

**Fahman Saeed[1], Muhammad Hussain[2]\*, Hatim A Aboalsamh[2], Fadwa Al Adel[3], Adi Mohammed Al Owaifeer[4]**

1. The authors Fahman Saeed is with Department of Computer Science, King Saud University and Imam Mohammad Ibn Saud Islamic University (IMSIU), Riyadh, Saudi Arabia. (emails: (faesaeed@imamu.edu.sa)

2. The authors Muhammad Hussain, and Hatim A Aboalsamh are with .Department of Computer Science, King Saud University, Riyadh, Saudi Arabia.

3. The author, Fadwa Al Adel is with Department of Ophthalmology, College of medicine, Princess Nourah bint Abdulrahman University.

4. The author Adi Mohammed Al Owaifeer is with Ophthalmology Unit, Department of Surgery, College of Medicine, King Faisal University, Al-Ahsa, Saudi Arabia, and also with King Khaled Eye Specialist Hospital, Riyadh, Saudi Arabia

\* *Correspondence:* Muhammad Hussain: mhussain@ksu.edu.sa

**Abstract:** Diabetic retinopathy (DR) is a leading cause of blindness in middle-aged diabetic patients. Regular DR screening with fundus images helps detect complications and slows disease advancement. Because manual screening takes time and is subjective, deep learning have been used to help graders. Pre-trained or brute force CNN models creation are used in existing DR grading CNN-based methods, not tailored to fundus image complexity. To solve this problem, we present an approach for automatically customizing CNN models based on fundus image lesions. It uses k-medoid clustering, Principal component analysis(PCA), and inter-class and intra-class variations to determine the CNN model's depth and width. The designed models are lightweight, adapted to the internal structures of fundus images, and encode the discriminative patterns of DR lesions. The technique is validated on a local dataset from King Saud University Medical City, Saudi Arabia and two challenging Kaggle datasets: EyePACS and APTOS2019. The auto-designed models outperform well-known pre-trained CNN models such as ResNet152, Densnet121, and ResNeSt50, as well as Google's AutoML and auto-keras models that are based on neural architecture search (NAS). The proposed method outperforms current CNN-based DR screening methods. The proposed method can be used in a variety of clinical settings to screen for DR and refer patients to ophthalmologists for further evaluation and treatment.

## 1. Introduction

Diabetes is a leading global health dilemma. One of its serious complications is diabetic retinopathy (DR), which has a prevalence of 34.6% worldwide and is considered a primary cause of blindness among middle-aged diabetic patients [1, 2]. A patient has a high DR risk if he or she has diabetes for a long time or is poorly managed. The DR treatment at its early stage slows down the retinal microvascular degeneration process. Graders manually screen fundus images for detecting DR prognosis, which is time-consuming and subjective [3-5]. On the other hand, screening a large number of diabetic patients for the possible prevalence of DR puts a heavy load on graders and reduces their efficiency. It necessitates intelligent systems for DR screening, and many ML-based systems have been proposed that show good results on public data sets. However, their performance is not certain in real DR screening programs, where there are different ethnicities, and the retinal fundus images are captured using different cameras. These factors affect these systems' performance and remain a challenge in their widespread use [6].

Deep CNN has shown remarkable results in many applications [7-11] and has been employed for DR screening [2, 12-14]. A CNN model usually involves a large number of parameters and needs a large amount of data for training. A brute force approach, which has been widely used for DR screening, is to adopt a highly complex CNN model designed for object recognition and pre-trained on the ImageNet dataset, and fine-tune them using fundus images [15-17]. As the ImageNet dataset consists of natural images, and the structural patterns of natural images and fundus images are entirely different, the architectures of the fine-tuned models do not adequately encode the fundus images. In addition, the complexity of pre-trained models is very high and not customized to DR screening from fundus images.

Instead, CNN models are manually designed from scratch. The design process starts with a CONV layer of a small width (i.e., the number of filters) and increases the widths of CONV layers by a fixed ratio as the network goes deeper [15-17]. There is no way to know what should be the depth (i.e., the number of layers) of a CNN model; a hit-trial strategy is used to fix the depth. In addition, CNN models are trained using iterative optimization algorithms such as stochastic gradient descent algorithms, and their convergence heavily depends on the initial guess of learnable parameters. Different data-independent [18, 19] and data-dependent [20, 21] approaches have been proposed to initialize them.

Alternatively, automated Machine Learning (AutoML) has developed into a significant area of research due to the widespread application of machine learning techniques [22]. AutoML's purpose is to make machine learning models accessible to those with limited machine learning prior knowledge. Some of the most commonly used methods for employing machine learning (ML) are easily available and may be used with just one or two lines of code. These systems include Auto-WEKA, Hyperopt-Sklearn, TPOT, Auto-Sklearn, and Auto-Keras [23-30]. Efforts have been made in automating model selection and tuning hyperparameters automatically, and so forth. Within the perspective of profound NAS stands for learning, neural architecture search [31], which aims to determine the optimal neural network architecture for a given learning task and dataset, has evolved into a highly effective computational tool for AutoML[32, 33]. It achieved competitive performance on the CIFAR-10 and Penn Treebank benchmarks by utilizing a reinforcement learning-based search strategy; consequently, NAS became a mainstream research topic in the machine learning community. NAS is prohibitively expensive and time consuming in terms of computation [34]. Zoph and and Le in [31] utilize massive computational resources (800 GPUs for three to four weeks) to achieve their result.

The preceding discussion demonstrates that developing an AutoML-customized lightweight CNN model for DR screening that uses a small subset of the target dataset and consumes less resources in a variety of clinical settings is difficult; it entails answering three design questions: (i) what must be the depth of the model, (ii) what must be the width of each of its convolutional (CONV) layer, i.e., the number of its kernels, and (iii) how to initialize the learnable parameters. To address these questions, we propose a constructive data-dependent approach for designing CNN models for DR screening under diverse clinical settings that automatically determines the depth of the model, the width of each CONV layer and initializes the learnable parameters. A custom-designed model takes a fundus image as input and grades it into normal or DR levels. We validated the proposed approach on three datasets: a local DR dataset from King Saud University Medical City, Saudi Arabia, two benchmark Kaggle datasets: EyePACS [35] and APTOS2019 [36]. Specifically, the main contributions of the paper are as follows:

- We proposed a constructive data-dependent AutoML approach to design lightweight CNN models customized to DR screening under various clinical settings. It automatically determines the depth of the model, the width of each COVN layer and initializes the learnable parameters using the fundus images dataset.

- To corroborate the usefulness of the proposed approach, we applied it to build an AutoML custom-designed lightweight CNN architectures for three datasets.
- We performed extensive experiments to show that the custom-designed lightweight CNN models compete well with the pre-trained models such as ResNet [16], DenseNet [17], ResNeSt [37], an AutoML NAS method, and other state-of-the-art methods for DR screening.

The layout of the rest of the paper is as follows: the literature view is presented in Section II, datasets are described in Section III, the detail of the proposed method is given in Section IV, the detail of experiments and the results are presented in Section V, and finally, Section VI concludes the paper.

## 2. PREVIOUS WORK

Different methods have been introduced for automatic DR screening; an extensive literature review is given in [38-41]. There are some efforts to compress and reduce the complexity of existing pre-trained CNN models by weights pruning [42, 43] or filters pruning [44-46]. First, we provide an overview of the previous work on building a deep model and initializing its weights and then give an overview of the state-of-the-art techniques for DR diagnosis.

**a) Data-dependent and auto deep models**

Different researchers employed principal component analysis (PCA) in various ways to build deep networks. Chan et al. [47] created an unsupervised two-layer model (PCANet). It is not an end-to-end model and is used only for feature extraction. Philipp et al. [20] used PCA to re-initialize pre-trained CNN models to avoid vanishing or exploding gradient problems. Suau et al. [21] used PCA and correlation to compress the filters of pre-trained CNN models. Seuret et al. [48] employed PCA to initialize the layers of stacked auto-encoders (SAEs). The above PCA-based methods have been employed for designing a CNN-like model for feature extraction, data-dependent re-initialization of the pre-trained models, or compressing their weights to reduce their complexity, but not for the data-dependent design of end-to-end CNN models.

Zhong et al. [49] introduced a method to build a BlockQNN module automatically using the block-wise setup, Q-Learning paradigm and epsilon-greedy exploration, and stack them to get the automatic CNN model. They evaluated their method using CIFAR-10, CIFAR-100, and ImageNet. It needs a lot of computational resources. They used 32 GPUs and got the best CNN model with BlockQNN after three days and Faster BlockQNN after 20 hours.

AutoML's initial effort was led by academia and machine learning practitioners, followed by startups. Auto-Weka (2013) [50] from the Universities of British Columbia (UBC). Following that, the University of Freiburg published Auto-sklearn (2014) [51]. The University of Pennsylvania developed TPOT [25](2015). After succeed of Zoph and and Le in [31] to performed comparably to the CIFAR-10 and Penn Treebank benchmarks, numerous recent efforts to develop NAS [52, 53] incorporate modern design elements previously associated with hand-crafted architectures, such as skip connections, which enable the construction of complex, multi-branch networks. To maximize efficiency, state-of-the-art systems employ cell-search spaces [54], which involves configuring only repeated cell-architectures rather than the global architecture, and employ gradient-based optimization [55]. Since 2013, Bayesian optimization has achieved several early successes in NAS, resulting in state-of-the-art vision architectures [56]. Google cloud AutoML based on NAS method is one of the famous auto deep learning models generating [57]. It utilizes transfer learning and neural architecture search (NAS) to determine the optimal network architecture and hyper-parameter configuration for that architecture that minimizes the model's loss function [58]. Another method for autoML-based NAS for generating deep learning models is auto-Keras (2017) [27] from Texas A&M University, which runs on top of Keras, Tensorflow, and Scikit-learn

b) **DR Screening methods**

Clinical DR screening categorizes a patient based on fundus images into different grades: level 0 (normal), level 1(mild), level 2 (moderate), level 3 (severe), and level 4 (proliferative). In the state-of-the-art on DR screening, various deep learning-based methods have been proposed, which address mainly three image-level DR grading scenarios: (i) scenario 1(SC1): normal and different levels of DR severity – a multi-class problem (ii) scenario 2(SC2): normal (level 0) vs. DR (levels 1~4) – a two-class problem, (iii) scenario 3(SC3): non-referral (level 0 and 1) vs. referral (levels 2-4) – a two-class problem. In the following paragraphs, we give an overview of the state-of-the-art best methods. Islam et al. [59] built a hand-designed CNN model consisting of 18 layers and 8.9 million learnable parameters. Its evaluation on the EyePACS dataset gave a sensitivity of 94.5%, a specificity of 90.2% for SC2, and a sensitivity of (98%) and a specificity of (94%) for SC3. Li et al. [60] introduced two hand-designed CNN models with 11 and 14 layers for feature extraction from the EyePACS dataset. The features from both models are fused and classified using an SVM classifier. They achieved an accuracy of 86.17% for SC1 and an accuracy of 91.05%, a sensitivity of 89.30%, and a specificity of 90.89% for SC2 using 5-fold cross-validation. Challa et al. [61] built a CNN model consisting of 10 layers for the EyePACS dataset and obtained an accuracy of 86% for SC1. Tymchenko et al. [62] built an ensemble of 20 CNN models. The ensemble used five versions of each of four pre-trained models: SE-ResNetXt50 with input sizes of 380x380 and 512x512, EfficientNet-B4, and EfficientNet-b5. It was fine-tuned using the APTOS2019 dataset. They got an accuracy of 91.9%, a sensitivity of 84%, specificity of 98.1%, and Kappa of 96.9% for SC1 on the APTOS2019 dataset. Sikder et al. [63] used an ensemble learning algorithm called ET classifier to classify the colored information of the fundus images from the APTOS2019 dataset. They filtered the dataset by removing many noisy samples and achieved an accuracy of 91% and a recall of 89.43% for SC1.

The above overview of the state-f-the-art methods shows that some methods used hand-designed CNN models and others employed pre-trained models and fine-tuning. For creating hand-designed models, the architectures of CNN models were fixed empirically using the hit-and-trial approach. In the case of fine-tuning, the complexity of the pre-trained models is very high and is not customized to the structures of fundus images.

## 3. Materials

We developed and validated custom-designed CNN models using two Kaggle challenge datasets: EyePACS [35] and APTOS2019 [36] and one local dataset collected at King Saud University Medical City (KSU-DR). Each dataset was preprocessed and augmented using the procedure described in Subsection 4.2.1. KSU-DR and EyePACS were divided into training (80%), validation (10%), and testing (10%). APTOS2019 consists of two sets: public training and public testing; 90% of the public training data was used for training, and the remaining 10% for validation and public testing for testing.

a) **KSU-DR**

The data was collected after getting approval from the local Institutional Review Board committee of King Saud University Medical City. The samples were collected randomly from fundus images of diabetic patients acquired during their routine endocrinologist's appointment at the funduscopic screening clinic. Fundus images were captured with a non-mydriatic fundus camera (3D-OCT-1-Maestro non-mydriasis fundus camera); a 45-degree fundus photo was captured from each eye. All patients were from Saudi Arabia, 44% were males, and 56% were females. The mean patient age was 53 years; 17% had type 1 diabetes, and 83% had type 2 diabetes. The mean duration of diabetes was 18 years (ranging from 4 - 42 years). Random samples of 1750 images were selected and graded by two expert ophthalmologists; 1024 were graded as normal, 477 as mild non-proliferative DR, 222 as moderate non-proliferative DR, 20 as severe non-proliferative DR, and 7 as proliferative DR (pdf).

b) **EyePACS**

EyePACS [35] consists of 88,702 color retinal fundus images with varying resolutions up to about 3000×2000 pixels [61], collected from 44,351 subjects, but only 35,126 labeled images are available in the public domain; most of the researchers used this set for the proposal of new algorithms [59, 64]. We also used 35,126 labeled images to design and evaluate the custom-designed CNN model. The images are graded into normal and 4 DR classes – mild, moderate, severe, and proliferative.

c) **APTOS2019**

APTOS2019 dataset [36] was published by the Asia Pacific Tele-Ophthalmology Society on the Kaggle competition website. Clinical experts graded the images into normal and 4 DR levels (mild, moderate, severe, and proliferative). The public domain version of this database contains 3662 fundus images for training and 1928 fundus images for testing

## 4. PROPOSED METHOD

### a) *Problem Formulation*

The problem is to predict whether a patient is normal or suffering from DR (with different levels of severity) using his/her retinal fundus images. Formally, let $R^{h \times w \times 3}$ be the space of color retinal fundus images with resolution $h \times w$ and $P = \{1, 2, \ldots, C\}$ be the set of labels where $C$ is the number of classes, which represent different DR grades; in case of two grades (i.e., normal and DR), $C = 2$, such that $c = 1$ means normal and $c = 2$ stands for DR; when there are five grades, $C = 5$, and $c = 1, 2, 3, 4, 5$ are the labels for normal, mild, moderate, severe, proliferative DR, respectively. To predict the grade of a patient, we need to define a mapping $\phi: R^{h \times w \times 3} \rightarrow P$ that takes a fundus image $x \in R^{h \times w \times 3}$ and associates it to a label $c \in P$, i.e. $\phi(x) = c$. We model the mapping function $\phi$ using a custom-designed CNN model.

### b) *Custom-designed CNN model*

The main constituent layer of a CNN model is the CONV layer, and the widely adopted CNN models contain a large number of CONV layers, e.g., VGGNet [65] contains 13 CONV layers. The number of layers (depth) and the number of filters in each layer (width) are fixed manually, keeping in view ImageNet challenge dataset [66], without following any formal procedure. Retinal fundus images have complex small-scale structures, which form discriminative patterns and are entirely different from those of the natural images in the ImageNet dataset. We design an AutoML CNN model for the DR problem by drawing its architecture from the fundus images; we determine the depth of a model and the widths of its CONV layers in a customized way using the discriminative information in fundus images specific to different DR levels. In this direction, the first design decision is about specifying the search space and extracting discriminatory information. For this purpose, first, we reduced the search space and select the most representative fundus images from the available DR dataset using the K-medoids clustering algorithm [67] and then apply PCA [68] to determine the widths of CONV layers and initialize them. The next design decision is about the depth (i.e., the number of CONV layers). We control the depth using the ratio of the between-class scatter matrix $S_b$ to the within-class scatter matrix $S_w$. Finally, motivated by the design strategy of ResNet [16], we add global pooling layers that follow the last CONV layer, and their outputs are fused and fed directly to a softmax layer. These layers control the drastic increase in the number of learnable parameters (which cause overfitting). The design process is described in detail below, and its overview is shown in Figure 1.

#### i. Preprocessing

The retinal fundus images are usually not calibrated and are surrounded by a black area, as shown in Fig(a). To center the retina and remove the black area around it, firstly, the retina circle is cropped, and the background is removed using the method presented in

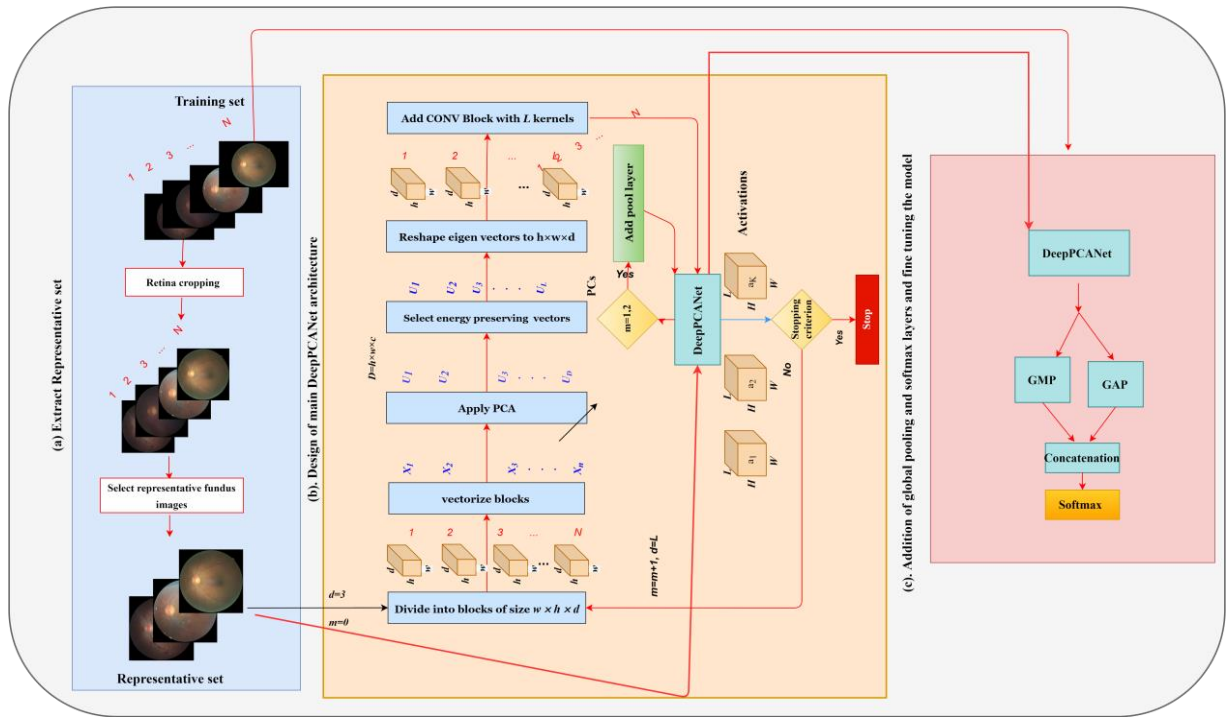

Fig. 1. Design procedure of DeepPCANet

[63], and then it is resized to 512 × 512 pixels. Usually, the DR datasets are imbalanced, i.e., the numbers of images of different classes are significantly different; we increase the data of minority

classes using data augmentation. We apply affine transformations to randomly rotate the image with an angle θ∈ (-180, 180).

*ii.* **Selection of representative fundus images**

We used three method (K-means [69] and K-medoids [67], and randome samples) for selecting the representative fundus images to drived the customize DeepPCANet model and test it to select the best one based on EyePACS training dataset. The discriminative features are extracted from training fundus images for clustering using the efficient LGDBP descriptor proposed in [70]. The number *K* of clusters for K-mean and K-medois is specified using the gap statistic method [71].

As indicated in Table 1, the K-medoids gave the best results and are the most precise. Due to the fact that K-means gives mean feature vectors as cluster centers, it is inadequate at selecting representative fundus images, and outliers are a serious concern. On the other hand, because the K-medoids algorithm selects representative fundus images as cluster centers, using representative fundus images is appropriate. Both K-medoids and K-mean outperform the random fundus image model.

TABLE 1. COMPARISON BETWEEN CLUSTERING METHODS BASED ON EYEPACS(SC1)

| Dataset | Model | ACC % | SE % |
|---|---|---|---|
| EyePACS | Random fundus images | 85.12 | 81.33 |
| | K-mean | 89.32 | 83.45 |
| | **K-medoids** | **94.22** | **86.56** |

### *iii.* Designing the main DeepPCANet architecture

The design of the AutoML customized architecture of DeepPCANet needs to address the two questions, i.e., (i) what should be the depth of the model and (ii) what should be the width of each CONV layer. These questions are addressed by an iterative algorithm that incrementally adds CONV layers and stops when a specific criterion is satisfied. It is based on the idea of exploiting discriminative information of fundus images to select the number of kernels in a CONV layer and initialize them. It takes representative fundus images $RI_j$, $j = 1, 2, 3, …, K$ as input, divides them into patches of size 7×7. The patches are vectorized and used to determine the number of kernels and initialize them. One possible idea is to cluster the patches and select the cluster centers as kernels, but the issue is choosing the number of clusters. We go for a simple and effective procedure, i.e., we employ PCA because it reduces the redundancy and causes to determine the kernels and their number, exploiting the discriminative information in the patches. The principal components (PCs), i.e., the eigenvectors along which the maximum energy is preserved, serve as kernels of the first layer. After computing the PCs, the DeepPCANet is initialized with an input

---

**Algorithm 1**: **To design the main DeepPCANet Architecture**

**Input**: Representative fundus images: $RI_j, j = 1, 2, 3, …, K$ of size $W \times H$ and the corresponding class labels $c = 1, 2, …, C$; Energy threshold $\varepsilon$

**Output**: The main architecture of DeepPCANet Architecture

**Processing**

**Step 1:** Initialize DeepPCANet with an input layer and set $w = 7, h = 7, d = 3, m = 0$ (*number of layers*)

**Step 2:** Set $a_j = RI_j$, $j = 1, 2, 3, …, K$, and *TRP (previous TR)* = *0*.

**Step 3:** Divide $a_j$, $j = 1, 2, 3, …, K$, into blocks $b_{ij}, i = 1, 2, 3, …, B$, $j = 1, 2, 3, …, K$, of size $w \times h \times d$, where $d$ is the number of channels (feature maps) in $a_j$ and $B$ is the number of blocks created from each $a_j$.

**Step 4:** Flatten $b_{ij}$ into vectors $x_i \in R^D, i = 1, 2, … M$, where $M = K \times B$, and $D = w \times h \times d$.

**Step 5:** Compute zero-center vectors $\phi_i$, $i = 1, 2, … M$ such that $\phi_i = x_i - \overline{x}$, where $\overline{x} = \frac{1}{M}\sum_{l=1}^{M} x_i$.

**Step 6:** Compute the covariance matrix $C = AA^T$, where $A = [\phi_1 \ \phi_2 \ … \ \phi_M]$.

Calculate the eigenvalues $\lambda_j$ and eigenvectors $u_j$ ($j = 1, 2, … D$) of the covariance matrix C.

**Step 7:** Select $L$ eigenvectors $u_i, i = 1, 2, …, L$ ($L < D$) corresponding to the $L$ largest eigenvalues such that $\frac{\sum_{l=1}^{L} \lambda_l}{\sum_{j=1}^{D} \lambda_j} \geq \varepsilon$, where $\varepsilon$ determines the level of energy to be preserved (e.g. $\varepsilon = 0.99$, for 99% energy preservation).

**Step 8:** All other eigenvectors corresponding to the $\frac{\sum_{l=1}^{L} \lambda_l}{\sum_{j=1}^{D} \lambda_j} < \varepsilon$ are summation as one eigenvector and stacked to the L eigenvectors

**Step 9:** Reshape $u_i, i = 1, 2, …, L + 1$ to kernels of size W×H×D, and add the CONV block to DeepPCANet; Update $m = m + 1$.

**Step 10:** If m = 1 or 2, add a max pool layer with a pooling window of size 2×2 and stride 2 to DeepPCANet.

**Step 11:** Compute the activations $a_j$, $j = 1, 2, 3, …, K$ of representative fundus images $RI_j$, $j = 1, 2, 3, …, K$ such that $a_j$ = DeepPCANet($RI_j$).

**Step 12:** Compute the ration $TR = \frac{Trace(Sb)}{Trace\ (Sw)}$ where $S_w = \sum_{i=1}^{C} \sum_{j=1}^{n_i}(x_j - \mu_i)(x_j - \mu_i)^T$ and $S_b = \sum_{i=1}^{C} n_i \ (\mu_i - \mu)(\mu_i - \mu)^t$.

**Step 13:** If $TRP\ (previous\ TR)\ \leq TR$, set $TRP = TR$, $W = 3, H = 3, D = L$, and go to Step 3, stop otherwise.

layer and a CONV block (BN+RELU+CONV) with kernels equal to the number of PCs; the kernels are initialized by reshaping the PCs. Please note that we fix the size of patches to 7×7 so that the size of kernels of the first CONV layers is 7×7 following the convention of most of the existing CNN models like Inception [15], ResNet [16], and DenseNet [17]. Using the current architecture of DeepPCANet, activations $a_j$, $j = 1, 2, 3, …, K$ of representative fundus images $RI_j$, $j = 1, 2, 3, …, K$ are calculated. Inspired by the Fisher ratio [72], using these activations, the ratio of the trace ($TR$) of between-class scatter matrix $S_b$ to the trace of within-class scatter matrix $S_w$ is calculated and is used to decide whether to stop or add another CONV block. The new CONV blocks continue to be added as long as TR continues to increase. This criterion ensures that the features generated by DeepPCANet have large inter-class variation and small intra-class scatter. To add a CONV block, the above procedure is repeated with activations $a_j$, $j = 1, 2, 3, …, K$. To reduce the size of feature maps for computational efficiency, pooling layers are added after the first and second CONV blocks. As the kernels and their number is determined from the fundus images, each layer can have a different number of filters. The detail of the design procedure is elaborated in Algorithm 1. It is to be noted that the PCs ($u_i$), which are used to specify the kernels of a CONV layer, are orthogonal and capture most of the variability in input fundus images, without redundancy, in the form of independent features. The PCs are selected so that the maximum energy is preserved. The energy is measured in terms of the corresponding eigenvalues i.e. $Energy = \frac{\sum_{l=1}^{L} \lambda_l}{\sum_{j=1}^{D} \lambda_j}$ [21, 73] and a threshold value is used to ensure that a certain percentage of energy (e.g., 99%) is preserved. The threshold value of 99% preserves the maximum energy with 209 (*L*) PCs for CONV1 in EyePACS dataset, as shown in Figure 2. The depth of AutoML CNN model and the width of each layer are important factors determining the model complexity. Step 7 of Algorithm 1 adaptively determines the best number of kernels that ensure the preservation of maximum energy of the input image. Step 9 initializes the kernels to be suitable for the DR domain. The selected kernels extract the features from fundus images (5 classes) so that the variability of the structures in fundus images is maximally preserved. It is also essential that the features must be discriminative, i.e., have large inter-class variance and small intra-class scatter as we go deeper in the network; it is ensured using the trace ratio $TR = \frac{Trace(Sb)}{Trace\ (Sw)}$, the larger the value of the trace ratio, the larger the inter-class variance, and the smaller the intra-class scatter [72]. Step 13 in Algorithm 1 allows adding CONV layers as long as $TR$ increases and determines the data-dependent depth of DeepPCANet. As shown in Figure 4, the maximum ratio is at layers 4, 5, and 16 for KSU-DR, APTOS2019, and EyePACS, respectively. It means that the suitable depth of the DeepPCANet model for the KSU-DR dataset is four layers (Figure 4(*a*)), for APTOS2019 is five layers ((Figure 4(*b*)), and for EyePACS dataset is 16 layers ((Figure 4(*c*)). The model for EyePACS is deeper because it contains many poor-quality fundus images, and there is the possibility of label noise because only one expert graded each image in this dataset. Each dataset was collected from a different region and under different conditions using different cameras, so the architecture of the DeepPCANet model is different for each dataset.

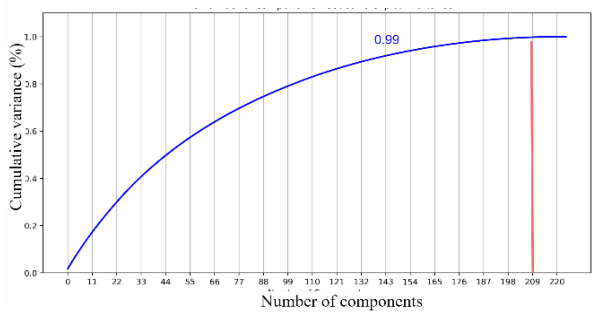

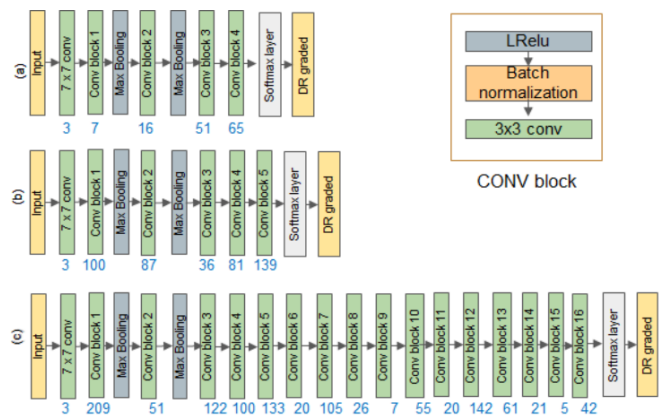

Fig. 2. Selecting the best threshold. The appropriate threshold is (0.99%) and the best number of eigenvectors is 209.

Fig. 3. DeepPCANet architecture for (a) APTOS (b) KSU-DR , and (c) EyePACS datasets

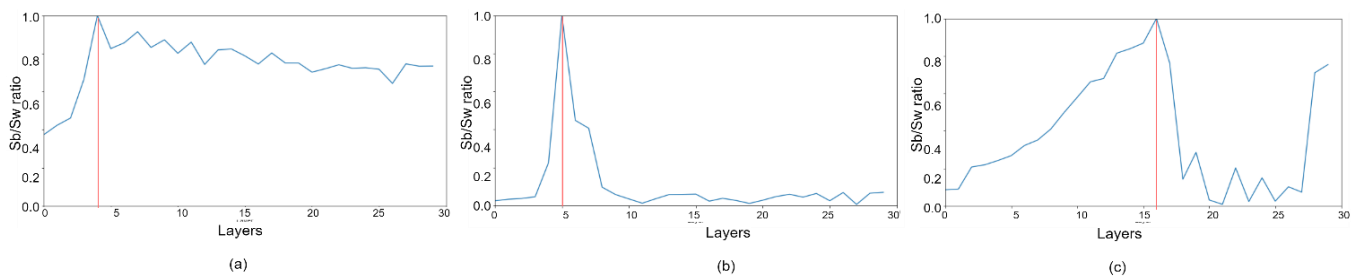

Fig. 4. Trace ratio of between class scatter and within class scatter. The depth for (a) APTOS2019 dataset is 4 layers, (b) KSU-DR dataset is 5 layers, and (c) EyePACS dataset is 16 layers.

    *iv.* **Addition of Global Pool and Softmax layers**

The dimension of the activation of the last CONV block is *W×H×L*. If it is flattened and passed to an FC layer, the number of learnable weights and biases of the FC becomes excessively large, which leads to overfitting. To overcome this issue, the activation of the last CONV block is passed simultaneously to global average pooling (GAP), and global max-pooling (GMP) layers [74], which extract the mean and largest feature from each feature map, and these features are fused using a concatenation layer. Both GAP and GMP help reduce the number of learnable parameters and extract discriminative features from the activation. Finally, a softmax layer is introduced as a classification layer, and the output of the concatenation layer is passed to this layer, as shown in Figure 1(c). A dropout layer is also added after the last CONV layer to overcome the overfitting problem.

    *v.* **Finetuning the DeepPCANet Model**

After determining the architecture of AutoML custom-designed DeepPCANet, it is fine-tuned using the training and validation sets. Fine-tuning involves various hyper-parameters: the optimization algorithm, learning rate, batch size, activation function, and dropout probability. We employed the Optuna optimization algorithm [75] to determine the best values of the hyperparameters. We tested three optimizers (Adam, SGD, and RMSprop), learning rate between 1e-5 and 1e-1, four batch sizes (5, 10, 15, 20), three activation functions (Relu, LRelu, and Sigmoid), and dropout probability between 0.25 and 0.50. After training for ten epochs, the Optuna returned the best hyperparameters for each dataset, as shown in Table 2. The number of kernels in each layer of each model is based on an energy threshold of 0.99. The models for APTOS2019, KSU-DR, EyePACS datasets (5 classes each) are DeepPCANet-4, DeepPCANet-5, and DeepPCANet-16, respectively, and their specifications are shown in Figure 3. Each dataset has different AutoML architecture because each one is from different ethnicities; EyePACS dataset is from USA, APTOS2019 is form India, and KSU-DR is from KSA; as well as the use of retinal images captured using different cameras. To confirm the distinct architectures for the three DR datasets, we combined the extracted K-medoids fundus images into a single dataset and generate a custom DeepPCANet and tested it on the three datasets. As illustrated in Table 3, the outcome is not as good as that obtained using the customized DeepPCANet for each DR dataset as illustrated in Tables 4 and 5. After fixing the hyperparameters, each model is fine-tuned using training and validation sets for 100 epochs. The fine-tuned model is tested using the testing set.

TABLE 2. THE BEST HYPERPARAMETERS FOUND USING OPTUNA ALGORITHM (SC1)

| Dataset | Activate function | learning rate | Patch's size | optimizer | dropout |
|---|---|---|---|---|---|
| **KSU-DR** | LRelu | 0.0001 | 10 | RMSprop | 0.50 |
| **EyePACS** | LRelu | 0.0055 | 10 | RMSprop | 0.38 |
| **APTOS2019** | LRelu | 0.0007 | 5 | RMSprop | 0.40 |

## 5. Experiments and results

This section first describes the evaluation protocol and the experiments performed to evaluate the proposed method and then presents the results.

### a) Evaluation Protocol

We determined the architecture of the DeepPCANet for each DR dataset and fine-tuned it using the training set of the corresponding DR database; the detail is given in Section 3. After that, the performance of each model was evaluated using the test set of the related database. To validate the usefulness and the superiority of the model design technique, we compared custom-designed models with the widely used state-of-the-art pre-trained CNN models like ResNet [16], DenseNet [17], and ResNeSt [37], which have shown outstanding performance for various computer vision applications. Also we compared it with AutoML models (Google cloud AutoML) and Auto-keras. We fine-tuned the competing models using the same procedure as was employed for DeepPCANet on each dataset.

For evaluation, we adopted three scenarios SC1 [60, 61], SC2 [59, 60], and SC3, as described in Section II.B. We evaluated the AutoML custom-designed models using SC1 and SC2 on APTOS2019 and EyePACS and SC3 in EyePACS. But evaluation on the KSU-DR dataset was done using SC2 because the number of images for five classes is not enough. In addition, we used four commonly used metrics in medical application and deep learning models: accuracy (ACC), sensitivity (SE), specificity (SP), and Kappa  [13, 76-79].

### i) Five Class Problem (SC1)

Using the APTOS2019 and EyePACS datasets, we built DeepPCANet-4 and DeepPCANet-16 models, respectively, for SC1 using the respective training sets and fined-tuned them using the corresponding training and validation sets (see detail in Section 3). After fine-tuning, the models were evaluated on test datasets of EyePACS and APTOS2019; the results are shown in Table 3. The results of the ResNet152, DensNet121, and ResNeSt50 models, fine-tuned using the same training set and evaluated using the same testing set as for from EyePACS and APTOS2019, are also shown in Table 3. The results show that DeepPCANet-4 and DeepPCANet-16 outperform RsenNet152, DesneNet121, and ResNeSt50 on both datasets in terms of all metrics; in particular, in both cases, the sensitivity and Cohen's Kappa are higher than those of RsenNet152, DesneNet12, and ResNeSt50, Cohen's Kappa is considered more robust statistical measure than accuracy [80, 81]. The DeepPCANet-4 has the lowest number of FLOPs (1.36 G) and learnable parameters ( 63.7 K) among all competing models, as shown in Table 3.

TABLE 3 THE BEST HYPERPARAMETERS FOUND USING OPTUNA ALGORITHM

| Dataset | Model | Performance (%) | | | |
|---|---|---|---|---|---|
| | | ACC | SE | SP | Kappa |
| **EyePACS(SC1)** | PCANet | 73 | 28 | 83 | 8 |
| **APTOS2019(SC1)** | model(Mixed | 88 | 36 | 91 | 51 |
| **KSU-DR(SC2)** | dataset) | 80 | 81 | 81 | 59 |

DeepPCANet-16 has fewer learnable parameters than the pertained ResNet152, DensNet121, and ResNeSt50 and also has less number of FLOPs than ResNet152 and Res-

**TABLE 4. COMPARISON BETWEEN DEEPPCANET MODELS AND THE PRETRAINED MODELS FOR SC1 SCENARIO. M AND K STAND FOR MILLION AND THOUSANDS**

| Dataset | Model | #FLOPs | # parameters | ACC % | SE % | SP % | Kappa % |
|---|---|---|---|---|---|---|---|
| APTOS2019 | ResNet152 | 5.6 M | 60.19 M | 95.25 | 88.22 | 96.97 | 88.15 |
| APTOS2019 | DensNet121 | 1.44 M | 7.98 M | 96.58 | 91.55 | 97.82 | 89.22 |
| APTOS2019 | ResNeSt50 | 5.39 M | 27.5 M | 97.11 | 92.29 | 98.2 | 90.82 |
| APTOS2019 | **DeepPCANet-4** | **1.36 M** | **63.7 K** | **98.21** | **95.29** | **98.9** | **94.32** |
| EyePACS | ResNet152 | 5.6 M | 60.19 M | 92.25 | 80.74 | 94.9 | 75.16 |
| EyePACS | DensNet121 | 1.44 M | 7.98 M | 91.14 | 80.07 | 95 | 74.84 |
| EyePACS | ResNeSt50 | 5.39 M | 27.5 M | 93.12 | 82.33 | 95.21 | 78 |
| EyePACS | **DeepPCANet-16** | **2.11 M** | **557.68 K** | **94.22** | **86.56** | **96.30** | **81.64** |

**TABLE 5 COMPARISON BETWEEN DEEPPCANET MODEL AND THE PRETRAINED MODELS FOR SC2 SCENARIO**

| Dataset | Model | #FLOPs | # parameters | ACC % | SE % | SP % | Kappa % |
|---|---|---|---|---|---|---|---|
| KSU-DR dataset | ResNet152 | 5.6 M | 60.19 M | 97.98 | 97.83 | 97.83 | 95.75 |
| KSU-DR dataset | DenseNet121 | 1.44 M | 7.98 M | 98.51 | 96.06 | 98.86 | 96.4 |
| KSU-DR dataset | ResNeSt50 | 5.39 M | 27.5 M | 99.47 | 99.46 | 99.46 | 98.93 |
| KSU-DR dataset | **DeepPCANet-5** | **1.375 M** | **73.66 K** | **99.5** | **99.5** | **99.5** | **98.99** |
| APTOS2019 | ResNet152 | 5.6 M | 60.19 M | 95 | 94.44 | 94.44 | 89.80 |
| APTOS2019 | DenseNet121 | 1.44 M | 7.98 M | 99.32 | 98.8 | 98.8 | 98.73 |
| APTOS2019 | ResNeSt50 | 5.39 M | 27.5 M | 98.33 | 96.54 | 96.53 | 94.22 |
| APTOS2019 | **DeepPCANet-4** | **1.36 M** | **63.7 K** | **99.7** | **99.44** | **99.44** | **99.3** |
| EyePACS | ResNet152 | 5.6 M | 60.19 M | 91.36 | 90.94 | 92.25 | 82.53 |
| EyePACS | DenseNet121 | 1.44 M | 7.98 M | 91.51 | 91.75 | 91.75 | 82.72 |
| EyePACS | ResNeSt50 | 5.39 M | 27.5 M | 90.53 | 90.92 | 90.92 | 79.04 |
| EyePACS | **DeepPCANet-16** | **2.11 M** | **557.68 K** | **94.44** | **94.28** | **94.28** | **88.71** |

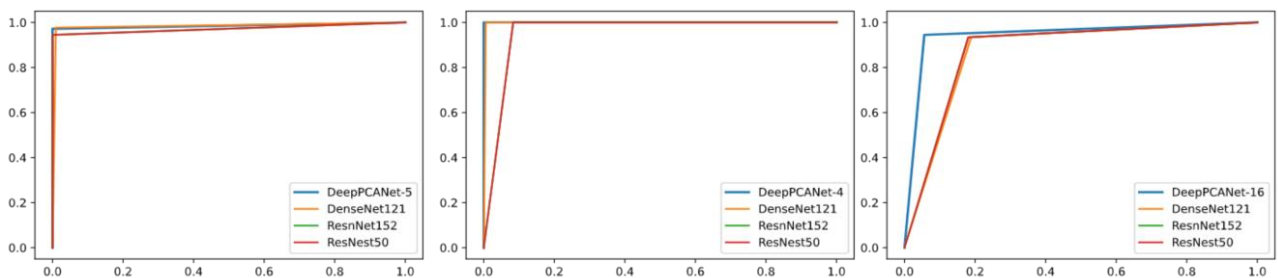

**Fig. 5.** ROC curve for custom-designed DeepPCANet and the pretrained models for SC2 scenario and datasets: (a) KSU-DR, (b) APTOS-2019, (c) EyePACS

NeSt50 models but slightly greater than DensNet121. In contrast, it has the best performance in terms of metrics on EyePACS dataset. ResNeSt50 has better performance than ResNet152 and DenseNet121. To compare the AutoML DeepPCANet to state-of-the-art AutoML methods, we use the most DR-intensive dataset available, EyePACS, based on scenario SC1. According to the NAS method [24], we test two AutoML methods; the Google cloud (vision) AtuoML [43], and Auto-keras [23]. We setup and generated the AutoML using Auto-keras methods locally using the same device and based on the representative set and finetuned the generated CNN model using training and validation set; then it is evaluated using test set as done with DeepPCANet-16. For Google cloud AutoML, we upload the representative, training, validation, and test sets to the Google cloud store and follow the same evaluation procedure. DeepPCANet-16 outperformed the Google cloud AutoML and Auto-Keras has less number FLOPs as shown in Table 6 but its performance is lower than both models. The FlOPs and number of parameters of Google cloud AutoML are hidden and it is shown only the precision(PR) and recall (SE) metrics. NAS algorithms are time-consuming and resource-intensive; they typically look for the cell structure, including the topology of the connections and the operation (transformation) that connects each cell. After that, the resulting cell is replicated in order to construct the neural network [82]. We used a basic simple cell structure throughout our AutoML DeepPCANet (LReLU, batch normalization layer, and CONV layer). The filters in the CONV are derived automatically from fundus' lesions and require less time. They optimized both the search architecture and hyper-parameters in NAS algorithms, whereas we derived the optimal DeepPCANet architecture firstly and then used Optuna to optimize the hyper-parameters as shown in Table 2.

TABLE 6. COMPARISON BETWEEN DEEPPCANET-16 AND ATUOML METHODS

| Dataset | Model | #FLOPs | # parameters | ACC % | SE % | PR % |
| --- | --- | --- | --- | --- | --- | --- |
| EyePACS | Auto-keras | 0.31 M | 15 M | 73 | 73 | 53 |
| | Google-AutoML | Hidden | hidden | -- | 71.43 | 79.1 |
| | **DeepPCANet-16** | 2.11 M | **557.68 K** | **94.44** | **94.28** | **96.12** |

*ii) Two Class Problem (SC2)*

We validated the DeepPCANet models' performance using the three datasets for SC2. The custom-designed models DeepPCANet-5, DeepPCANet-4, and DeepPCANet-16 for KSU-DR, APTOS2019, and EyePACS, respectively, which were designed and fine-tuned using only fundus images, outperform the highly complex CNN models such as RsenNet152, DesneNet121, and ResNeSt50, which were trained using ImageNet dataset and fine-tuned using fundus images, in terms of all metrics, as shown in Table 4. Though DenseNet121 outperforms ResNet152 and ResNeSt50 on the three datasets, its performance is not better than the custom-designed models. DeepPCANet-5 involves 1.375G FLOPs, which is smaller than the number of FLOPs of ResNet152, DenseNet121, and ResNeSt50. The number of learnable parameters of DeepPCANet-5 is 73.66K witch is much smaller than those of the pre-trained models ResNet152 (60.19 G), DenseNet121 (7.98 G), and ResNeSt50(27.5 G). In Figure 5, we provide illustrations of the ROC curves on the three datasets using the four models (customized DeepPCANet, ResNet152, DenseNet121, and ResNeSt50). It indicates that DeepPCANet models' performance is better than the three pre-trained models on the three datasets.

**b) Visualization**

To understand the decision-making mechanism of the custom-designed CNN models, we created the visual feature maps using the gradient-weighted class activation mapping (GradCam) visualization method [83]. The visual feature maps of two random fundus images generated by the DeepPCANet-5 model customized for the local KSU-DR dataset are shown in Figures 6(d) and 5(i).   The same fundus images were given blindly to two expert ophthalmologists at King Khalid Hospital of KSU, and they independently specified the lesion regions manually. Though there is a slight difference in the annotations of both experts, they agreed on most of the lesions, as shown in Figure 6 (b, c) for the fundus image (Figure 6a) from class moderate and Figure 6(g, h) for the fundus image (Figure 6f) from class pdf. The visual features maps of the DeepPCANet-5 model highlight the lesions annotated by both experts, as shown in Figure5(d, i). The yellow and orange splatter in Figure.i indicates that the DeepPCANet-5 model takes decisions based on the features learned from the lesion regions.

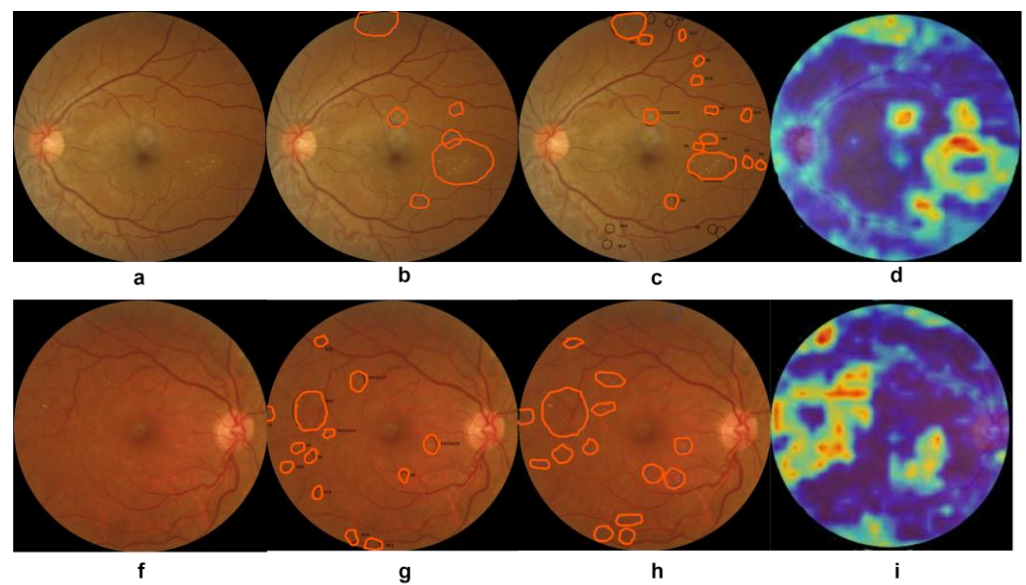

Fig. 6. Visualization of the decision making mechanism of DeepPCANet-5   model. (a) fundus image from class moderate, (b, c) lesions specified by experts 1 and 2, respectively, (d) DeepPCANet-5 map (f) fundus image from class pdf, (g,h) lesions specified by experts 1 and 2, respectively, (i) DeepPCANet-5 map.

## 6. Discussions

This study proposed a technique to auto custom-design a DeepPCANet model for a target DR dataset. The depth of the model and the width of each layer is not specified randomly or by exhaustive experiments. The custom-designed DeepPCANet models for DR screening have small depths and varying widths of CONV layers and involve a small number of learnable parameters. The results of the AutoML DeepPCANet models customized for the KSU-DR, APTOS2019, and EyePACS datasets (presented in Tables 4, 5, and 6) demonstrate that it outperforms the well-known highly complex pre-trained models ResNet152, DenseNet121, and ResNeSt50, as well as AutoML from Google and Auto - keras that were fine-tuned using the same DR datasets. Generally, the DeepPCANet got competitive performance with a small number of layers and parameters. As shown in Table 4, the custom-designed DeepPCANet models for the three datasets have a small number of parameters in thousands against that number in millions of ResNet152,

DensNet121, and ResNeSt50. DeepPCANet-4 and DeepPCANet-5 have fewer FLOPs than all pre-trained models and have better performance. The DeepPCANet-16 has fewer FLOPs than that of ResNet152 and ResNeSt50 and also has better performance. Though DensNet121 has fewer FLOPs than DeepPCANet-16, it has less performance and a large number of parameters. The reason for the lightweight structures and superior performance of custom-designed DeepPCANet models is that their architectures have been directly drawn from the fundus images, unlike the state-of-the-art CNN models, which have been mainly designed for object detection. In addition to comparing the custom-designed DeepPCANet models with famous pre-trained models, it is essential to validate their effectiveness in DR screening by comparing them to the state-of-the-art methods on two challenging datasets (APOTS2019 and EyePACS). DeepPCANet-4 generated for SC1 on the APTOS2019 dataset outperforms the state-of-the-art methods on the same dataset in terms of accuracy, sensitivity, specificity, and Kappa, as shown in Table 7. Though the Method by Tymchenko et al. 2020 [62] outperforms the DeepPCANet-4 in the Kappa score for the five-class problem (SC1), it has less accuracy, sensitivity, specificity, and it is based on a highly complex ensemble of 20 CNN models. For the same scenario, the DeepPCANet-16 designed for EyePACS outperforms the existing methods in accuracy and specificity. The method by Islam et al. 2018 [59] got higher sensitivity, but their model is more complex, and it was tested on 4% of the dataset, as shown in Table 7. For the SC2 (normal vs. DR levels), DeepPCANet-4 outperforms the method by Tymchenko et al. [62] in all metrics on APTOS2019. In this scenario, on the EyePACS dataset, as shown Table 7, the DeepPCANet-16 is better than other methods in accuracy, sensitivity, and specificity; the method by Islam et al. 2018 [59] is slightly better than DeepPCANet-16 in sensitivity, but it was tested only on 4% of the EyePACS dataset. The method of Chetoui et al 2020 [86] is better than DeepPCANet-16 wherease they used transfere learning based on Inception-Resnet-v2 witch has a high complexity and number of parameters. It consists of five convolutional layers, each of which is followed by batch normalization, two pooling layers, 43 inception modules, three residual connections, and the pooling of global averages and the use of two fully connected layers in conjunction with the rectified linear unit (ReLU). Wherase DeepPCANet-16 is a 16-layer structure that employs the basic CONV setup. The DeepPCANet-16, based on the EyePACS dataset for the SC3 (0 and 1 vs. DR levels), got less accuracy than Colas et al. [64] and Islam et al. [59] but got higher sensitivity and specificity, which are more important and robust than accuracy in the medical applications [87].

Table 7. Comparison between DeepPCANet models and state-of-the-art methods

| Paper | Method | Dataset | Performance (%) | | | |
|---|---|---|---|---|---|---|
| | | | ACC | SE | SP | Kappa |
| **Five classes (SC1)** | | | | | | |
| Sikder et al. 2019 [63] | Colored features extraction using En-sample | APTOS2019 | 91 | 89.54 | | |
| Tymchenko et al. 2020 [62] | An ensembled models with 3 CNN architectures fficientNet-B4, EfficientNet-B5, and SE-ResNeXt50 | APTOS2019 | 91.9 | 84 | 98 | 96.9 |
| Shorfuzzaman et al 2021 [84] | CNN-based transfer learning ensemble | aptos2019 (SC1) | 96.2 | 94.00 | - | - |
| **DeepPCANet-4** | **DeepPCANet model customized for APTOS2019 dataset** | **APTOS2019 (test: 40 % of public dataset)** | **98.21** | **95.28** | **98.85** | **94.32** |
| Lahmar et al 2021 [85] | Transfer learning (MobileNet V2) | APTOS2019(SC1) | 93.09 | 89.27 | 92.69 | - |
| Islam et al. 2018 [59] | A CNN model consisting of 18 layers with 3x3 and 4x4 kernels | EyePACS (test: 4% of dataset) | | 94.5 | 90.2 | |
| Li et al. 2019 [60] | deep learning model based on DCNN | EyePACS (test: 10% of dataset) | 86.17 | | | |
| **DeepPCANet-16** | **DeepPCANet model customized for EyePACS dataset** | **EyePACS (test: 10 % of public dataset)** | **94.22** | **89.56** | **96.30** | **81.64** |
| **Normal vs DR (all DR levels) (two classes) (SC2)** | | | | | | |
| Tymchenko et al. 2020 [62] | Ensembled models with 3 CNN architectures fficientNet-B4, EfficientNet-B5, and SE-ResNeXt50 | APTOS2019 | 99.3 | 99.3 | 99.3 | 98.6 |
| **DeepPCANet-4** | **DeepPCANet model customized for APTOS2019 dataset** | **APTOS2019(test: 10 % of public dataset)** | **99.7** | **99.44** | **99.44** | **99.3** |
| Islam et al. 2018 [59] | A CNN model consisting of 18 layers with 3x3 and 4x4 kernels | EyePACS (test: 4% of public dataset) | | 94.5 | 90.2 | |
| Li et al. 2019 [60] | Features extraction using deep learning model based on DCNN and SVM classification | EyePACS | 91.05 | 89.30 | 90.89 | |
| Chetoui et al 2020 [86] | pretrained Inception-Resnet-v2 DCNN | EyePACS (SC2) | 97.9 | 95.8 | 97.1 | 98.6 |
| **DeepPCANet-16** | DeepPCANet model customized for EyePACS dataset | EyePACS (test: 10 % of public dataset) | 94.44 | 94.28 | 94.28 | 88.71 |
| **Non-referral (Normal & DR grad 1) vs referral (DR grade 2 to highest grade) (two classes) (SC3)** | | | | | | |
| Colas et al. 2016 [64] | A CNN model. End to end training. | EyePACS (train:89%, test:11%) | 96.2 | 66.6 | 94.6 | |
| Islam et al. 2018 [59] | A CNN model | EyePACS (train:96%, test:4%) | 98 | 94 | | |
| **DeepPCANet-16** | DeepPCANet model derived from EyePACS dataset | EyePACS (test: 10 % of public dataset) | 94.59 | 94.86 | 94.87 | 89.02 |

## 7. Conclusion

We introduced an approach to building an AutoML data-dependent CNN model (DeepPCANet) customized for DR screening automatically. This approach tackles the limitations of the available annotated DR datasets and the problem of a vast search space and huge number of parameters in a deep CNN model. It builds an auto lightweight CNN model customized for a target DR dataset using k-medoid clustering, principal component analysis (PCA) and inter-class and intra-class variations. The DeepPCANet model is data-dependent, and each DR dataset has its appropriate AutoML architecture. The customized models, DeepPCANet-5 for the local KSU-DR dataset, DeepPCANet-4 for APTOS2019, and DeepPCANet-16 for EyePACS dataset, outperform the pre-trained very deep and highly complex ResNet152, DenseNet121, and ResNeSt50 models fine-tuned using the same datasets and procedure. The performance, complexity, and number of parameters of the customized DeepPCANet models are less than ResNet152 and ResNeSt50 significantly. Though DensNet121 has fewer FLOPs than DeepPCANet-16, it has less performance and a large number of parameters. On the EyePACS dataset, compared to the Google cloud AutoML, and Auto-Keras, DeepPCANet-16 based on SC1 got better performance with less number of parameters. Using EyePACS dataset DeepPCANet-16 also compared to the state-of-the-art methods (for SC2 and SC3), the DeepPCANet-16 has less complexity and parameters and got a competitive performance. The DeepPCANet fails to predict DR grade from fundus images, which have poor quality. It could not grade some poor quality fundus images from the EyePACS dataset correctly; each image in this dataset was graded by only one expert from the geographic region of California, which can potentially lead to annotation bias. How the DeepPCANet can reliably predict the DR grade from poor quality fundus images is a subject of future work. Also, how the DeepPCANet can be generalized with different fundus datasets is a subject of future work.

**Declaration**

There is no conflict of interest.